\documentclass[preprint,5p]{elsarticle} 
\usepackage{amsmath}
\usepackage[export]{adjustbox}
\usepackage{graphicx}
\biboptions{sort&compress}

\title{Dependence of the kinetic energy absorption capacity of bistable mechanical metamaterials on impactor mass and velocity}

\author[1]{R. Fancher}
\author[1]{I. Frankel}
\author[4]{K. Chin}
\author[5]{M. Abi Ghanem}
\author[1]{B. MacNider}
\author[3]{L. S. Shannahan}
\author[3]{J. F. Berry}
\author[3]{M. Fermen-Coker}
\author[4]{A. J. Boydston}
\author[1,2]{N. Boechler}

\affiliation[1]{organization={Department of Mechanical and Aerospace Engineering, University of California, San Diego}, 
city={La Jolla},
state={CA},
country={US}}

\affiliation[4]{organization={Department of Chemistry, University of Wisconsin at Madison}, 
city={Madison},
state={WI},
country={US}}

\affiliation[5]{organization={Institut Lumière Matière, niversité Claude Bernard Lyon 1, CNRS}, 
city={Villeurbanne},
state={F-69622},
country={France}}

\affiliation[3]{organization={Army Research Laboratory}, 
city={Aberdeen Proving Ground},
state={MD},
country={US}}
            
\affiliation[2]{organization={Materials Science and Engineering, University of California, San Diego}, 
city={La Jolla},
state={CA},
country={US}}

\begin{document}

\begin{frontmatter}

\begin{abstract}

Using an alternative mechanism to dissipation or scattering, bistable structures and mechanical metamaterials have shown promise for mitigating the detrimental effects of impact by reversibly locking energy into strained material. Herein, we extend prior works on impact absorption via bistable metamaterials to computationally explore the dependence of kinetic energy transmission on the velocity and mass of the impactor, with strain rates exceeding $10^2$ s$^{-1}$. We observe a large dependence on both impactor parameters, ranging from significantly better to worse performance than a comparative linear material. We then correlate the variability in performance to solitary wave formation in the system and give analytical estimates of idealized energy absorption capacity under dynamic loading. In addition, we find a significant dependence on damping accompanied by a qualitative difference in solitary wave propagation within the system. The complex dynamics revealed in this study offer potential future guidance for the application of bistable metamaterials to applications including human and engineered system shock and impact protection devices. 

\end{abstract}

\end{frontmatter}

\section{Introduction}

The local magnification of mechanical forces as a result of a dynamic collision (impact) and their detrimental effects on natural and engineered systems have been studied extensively \cite{schiehlen2017long,meyers1994dynamic,harris2002harris,nesterenko2013dynamics}. A relatively new approach for the mitigation of damage induced by impact is the use of bistable structures, which in contrast to the more ubiquitous mechanisms of dissipation and scattering \cite{lakes2009viscoelastic,bouferra2005study,li2022ballistic,mason1948energy}, reduces the effect of impact by reversibly ``locking'' some of the energy imparted by a shock or impact into the form of strain energy \cite{wang2017harnessing,CaoReview2021}. Both the performance and reversibility of bistable structures for impact mitigation are attractive, as the energy locking mechanism could be used in conjunction with other mechanisms \cite{Kang2022}, such as dissipation, and the structures can ostensibly be reset in a controllable fashion for reuse. 

In addition to studies of the impact absorption characteristics of single bistable structures \cite{wang2017harnessing,CaoReview2021}, more recently, the energy absorption properties of bistable structures configured into multi-unit-cell mechanical metamaterials has been explored \cite{CaoReview2021,shan2015multistable,frenzel2016tailored,Valdevit2017a,Valdevit2017b,wu2018multi,ha2018design,Mailen2021,Kang2022}. In such studies, the material response has been studied in either relatively low rate regimes \cite{shan2015multistable,frenzel2016tailored,Valdevit2017a,Valdevit2017b,wu2018multi,ha2018design,Mailen2021}, where in some cases the response was suggested to be rate independent \cite{shan2015multistable}, while in other cases the impact response was not related to the material's wave dynamics \cite{Kang2022}. This can be placed in contrast with other studies of nonlinear solitary, or ``transition'', wave propagation in similar materials, which show significant dependence on the system excitation \cite{deng2020nonlinear,nadkarni2016unidirectional,raney2016stable,dauxois2006physics,Kochmann2017,katz2019solitary}. A notable work that brings these two features (impact absorption via bistability and nonlinear waves) together is that of Ref. \cite{katz2018solitary}, which described the effect that the shape of the bistable potential has on energy trapping and solitary wave emission, however, they did not study the dependence of absorption performance on impact conditions. 

\begin{figure*}[t!]
\centering
  \includegraphics[width=6in]{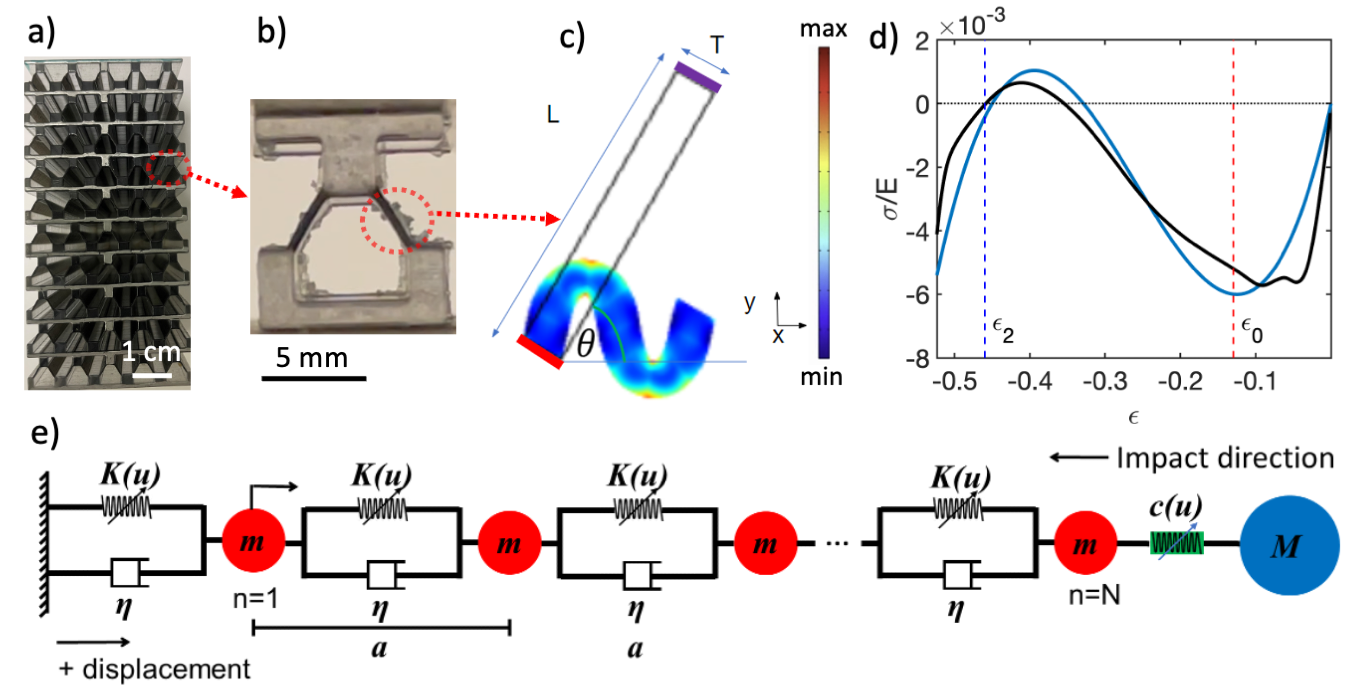}% 
\caption{Concept and modeling overview. (a) Photograph of a 3D printed model of the bistable mechanical metamaterial (with fewer unit cells than modeled herein). (b) Photograph of a 3D printed unit cell. (c) One beam in the unit cell design, showing the displacement of the beam near $\epsilon_2$ simulated using FEM, where the colorscale denotes the von Mises stress (arb.). (d) Non-dimensionalized effective stress and strain of the mechanical metamaterial (for $r=0.14$ and $\theta=60$ degrees), based upon experimental compression measurements (solid black line), and the 3rd order polynomial fit of the experimentally measured curve (solid blue line), which is used in the DEM model. The vertical dashed red and blue lines denote $\epsilon_0$ and $\epsilon_2$, respectively, and the horizontal dotted black line denotes zero stress. (e) Illustration of the DEM model.} 
\label{Modeling Concept}
\end{figure*}

In this work, we computationally study the dependence of kinetic energy transmission in a bistable mechanical metamaterial (shown in Fig.~\ref{Modeling Concept}) on the velocity and mass of the impactor, in wave dominated regimes (wavelengths less than the absorbing material's reference length) with maximum strain rates of approximately $195$ s$^{-1}$. Kinetic energy (KE) transmission is used as a performance metric herein as it has been previously shown to be closely related to damaging effects, for instance, in the case of behind armor blunt trauma \cite{national2011opportunities,sonden2009trauma}. The computational model of our mechanical metamaterial is a discrete element model (DEM) composed of a one-dimensional (1D) chain of masses connected by bistable nonlinear springs, with linear intersite damping, and a contact spring at the boundary that allows release of the impactor under tension during rebound. We find a large dependence of KE transmission on both impactor mass and velocity, ranging from significantly better to worse performance than a comparative linear material. We correlate said performance to solitary wave formation and give analytical estimates of idealized energy absorption capacity under dynamic loading. In addition, a significant effect is found from the inclusion of the intersite damping, accompanied by a qualitative difference in solitary wave propagation within the system. The complex dynamics revealed in this study offer potential future guidance for applications including improved packaging to prevent damage during shipment \cite{shan2015multistable}, personal protection equipment \cite{li2022ballistic,shan2015multistable,wang2015failure}, and crash mitigation for vehicles \cite{shan2015multistable,wu2018multi,sherwood1992constitutive}. 

\section{Description of metamaterial to be modeled}

The conceptual setup described herein is the impact of a relatively rigid ``impactor'' block onto an arbitrarily-sized ``absorbing'' material block. The dimensions of the absorbing block were chosen as $2d$ (height) by $d$ (both width and depth), with $d=5$ cm. Our designed absorbing material consists of a bistable mechanical metamaterial composed of a periodic array of structured unit cells based upon the geometry from Ref. \cite{shan2015multistable}. The unit cell of our bistable mechanical metamaterial is composed of two elastic beams surrounded by a lower-aspect ratio monolithic ``frame''. Figure \ref{Modeling Concept}(a,b) shows a 3D printed representation (with fewer unit cells and layers than modeled herein) of the bistable mechanical metamaterial, where the beams are composed of a rubber simulant material and the frame composed of an acrylic simulant. The thickness to length aspect ratio of the beam elements $r=T/L$ (where $T$ is the beam thickness and $L$ is its length) and the angle of the beam $\theta$ was chosen such that $r=0.14$ and $\theta=60$ degrees. Large deformation finite element method (FEM) simulations of beam deformation under compression are shown in Fig. \ref{Modeling Concept}(c), assuming a Neo-Hookean material model (Lam\'e Parameters: $\lambda=47.191$ GPa, $\mu=25$ GPa), fixed boundaries on the bottom end (red), with a prescribed y-direction displacement on the top end (purple, fixed in the x-direction), and free boundaries otherwise. We model the beam separately from the frame, because we approximate the frame itself to undergo minimal deformation as it is much thicker (structurally stiffer) than the beams. Furthermore, we assume that the two beams within the unit cell deform symmetrically with respect to each other. As such the mechanical response of a single beam can be extrapolated to determine the elasticity of the metamaterial. The choice of the beam aspect ratio was chosen to obtain moderately high energy locking without self-contact \cite{shan2015multistable}, but without too small of an energy barrier before unsnapping. The size of the frame was chosen to avoid self-contact while maintaining rigidity without excessive density per unit volume of the unit cell. Both the frame and geometry are natural candidates for future optimization studies. 

\section{Measurement of quasi-static bistable material response}

Figure \ref{Modeling Concept}(d) shows the experimentally measured mechanical response of a single 3D printed layer of bistable metamaterial (composed of the same material and geometry as in  Fig. \ref{Modeling Concept}(a,b), but less depth in the direction going into the page), along with a 3rd order polynomial fit of the measurement. The response of the layer was measured using a mechanical test frame in displacement control, where the rigid (acrylic simulant) top and bottom parts of the layer were attached to grips in a mechanical testing frame, allowing measurement of tension, as well as compression, following the snap through event. Herein a ``layer'' is defined as a row of unit cells where the normal is in the direction of the impactor velocity vector and along the long axis of the absorbing block. In Fig. \ref{Modeling Concept}(d), $\sigma$ and $\varepsilon$ are the effective bistable mechanical metamaterial stress and strain, respectively, and $E= 8.42e5$ N/m$^2$ is the measured small strain elastic modulus of the rubber simulant beam material. The transition to negative stiffness can be seen to occur in the fit at $\varepsilon_0 = -0.13$ and the second stability point at $\varepsilon_2 = -0.46$.

\section{Discrete element model}

Figure \ref{Modeling Concept}(e) shows a visualization of the DEM used to simulate the dynamics of the metamaterial undergoing impact. Each layer of unit cells in the absorbing material were described as lumped mass layers of mass $m$ connected by massless springs and dampers, with the impactor of mass $M$ interacting with the absorber via a contact spring. A DEM was chosen as a reasonable model due to the following key assumptions: i) That the model is designed to describe uniaxial loading of the impactor on the absorber block with minimal off axis loading effects; ii) near zero effective Poisson’s ratio of the lattice; and iii) The lattice is composed of stiffer and larger masses lumped within the material, separated by softer, lower mass elements. Regarding the this third assumption, more precisely, the vibrational frequencies of the separate mass and spring components must be much higher than the modal frequency of the two elements combined, where the spring deforms as the mass moves like a rigid body, such that the higher frequencies can be reasonably ignored. The sample mass was divided into equal “mass layers” in number equal to the amount of unit cells along the height of the material ($N$).  

A ``contact spring” with nonlinear stiffness, based on the Hertzian contact model \cite{hertz1881j}, between the top unit cell and the impactor mass was modeled to describe the impactor hitting the top of the sample and allowing the impactor to freely bounce rather than stick to the top of the lattice after impact. The equations of motion for the impactor particle (\textit{i.e.} particle $n=N+1$) is thus: 

 \begin{equation} \label{big_mass_v1}
M\ddot{y}_{N+1} = {C_{1}}([y_{N}-y_{N+1}]_+)^{3/2},
 \end{equation}
 
\noindent where the $[]_+$ denotes spring's inability to support tension (if the value in brackets is negative it equals zero), $y_n$ is the displacement of the $n$th layer, and $C_1$ is a fitting parameter (set to $-8.51e8$). As such, the contact stiffness (shown in Fig. \ref{Modeling Concept}(e)) is $c(u)=-\frac{3}{2} C_1 ([-u]_+)^{1/2}$, where $u$ is the spring stretch (positive in tension for both $u$ and $c(u)$). In addition to allowing the impactor to rebound, the use of a contact spring allowed for a better estimation of real impact conditions, including roughly describing the impactor coming into contact with the sample at a slight relative angle, or having asperities on the two surfaces. 

We then define the force-displacement relation of an individual layer: 

\begin{equation}
    \label{FLeqn}
    F_L(u)=\beta_{3}(u)^3 + \beta_{2}(u)^2 + \beta_{1}(u),
\end{equation}

\noindent where $F_L$ is positive in tension, $\beta_1= 2.257e4$ N/m, $\beta_2 = 1.187e7$ N/m, and $\beta_3 = 1.524e9$ N/m are coefficients from the fit of the  mechanical response measured for the 3D printed bistable layer (shown in Fig. \ref{Modeling Concept}(d)). The stiffness is thus defined as $K(u)=\partial F_L(u)/\partial u$ (shown in Fig.~\ref{Modeling Concept}(e)). Equation \ref{FLeqn} is used in all other elements of the DEM (corresponding to the metamaterial layers), where the top layer of the absorber ($n=N$) is: 

\begin{equation} \label{top_lattice_mass}
    \begin{split}
 {m}\ddot{y}_N &= {C}_{1}(y_N-y_{N+1})^{3/2}-F_L(y_{N}-y_{N-1}) \\ &-{\eta}(\dot{y}_{N} - \dot{y}_{N-1}),
    \end{split}
\end{equation}
 
\noindent where $\dot{y}_n$ and $\eta$ are the velocity of the $n$th particle and $\eta$ the damping coefficient, respectively. 

The equations of motion for layers from $n=2$ to $n=N-1$ are given by:
 
\begin{equation}\label{inbetween_masses}
    \begin{split}
{m}\ddot{y}_{n} =&F_L(y_{n+1}-y_{n})-F_L(y_{n}-y_{n-1})\\
+&{\eta}(\dot{y}_{n+1} - \dot{y}_{n}) -{\eta}(\dot{y}_{n} - \dot{y}_{n-1}).
\end{split}
\end{equation}

The equation of motion for the $n=1$ mass, next to the fixed boundary is then:
 
\begin{equation} \label{bottom_mass}
    \begin{split}
{m}\ddot{y}_{1} ={}& F_L(y_{2}-y_{1})-F_L(y_{1})\\
+&{\eta}(\dot{y}_{2} - \dot{y}_{1}) -{\eta}\dot{y}_{1}.
\end{split}
\end{equation}
 
The equations of motion were numerically integrated using the ODE45 integrator in MATLAB \cite{MATLAB:2019}, given an initial velocity $V$ applied to the impactor mass, to solve for particle displacements and velocities as a function of time. 

\section{Analytical estimate of nominal impact conditions}

The kinetic energy transmitted through the half-way-point of the absorbing block (with respect to the impactor velocity vector), or unit cell $n=N/2$, was chosen as the performance metric (so as to avoid boundary effects), which was then compared against that of a linearly coupled (with the exception of the contact spring) absorbing material. 

We then endeavored to estimate the impactor parameters to minimize KE transmission through $n=N/2$ of our chosen bistable mechanical metamaterial. The nominal impact velocity was estimated as:

\begin{equation} \label{v0eqn}
V_{0} = 2c_{0}\varepsilon_{0},
\end{equation}

\noindent where $c_{0}$ is the long wavelength linear soundspeed of the lattice \cite{meyers1994dynamic}. For all simulations, a unit cell length of $a=1$ mm was used. Equation \ref{v0eqn} is meant to estimate the minimum velocity threshold such that the snap through to the second stable state is induced. In an ideal scenario, the entire material between the $n=N/2$ and the impactor would then snap to the second stable state, and have locked the entire KE from the impactor into stored potential energy (PE). The PE absorbed from the first half of the absorber is then: 

\begin{equation} \label{PEeqn}
\text{PE} = \frac{N}{2}\int_{o}^{a\epsilon_2} F_{L}(u)du. 
\end{equation}
\noindent The nominal impactor mass $M_0$ was solved as the only remaining unknown when impactor KE (\textit{i.e.} $MV^2/2$) was set equal to PE.  

\begin{figure*}[h!]
\centering
  \includegraphics[width=\textwidth]{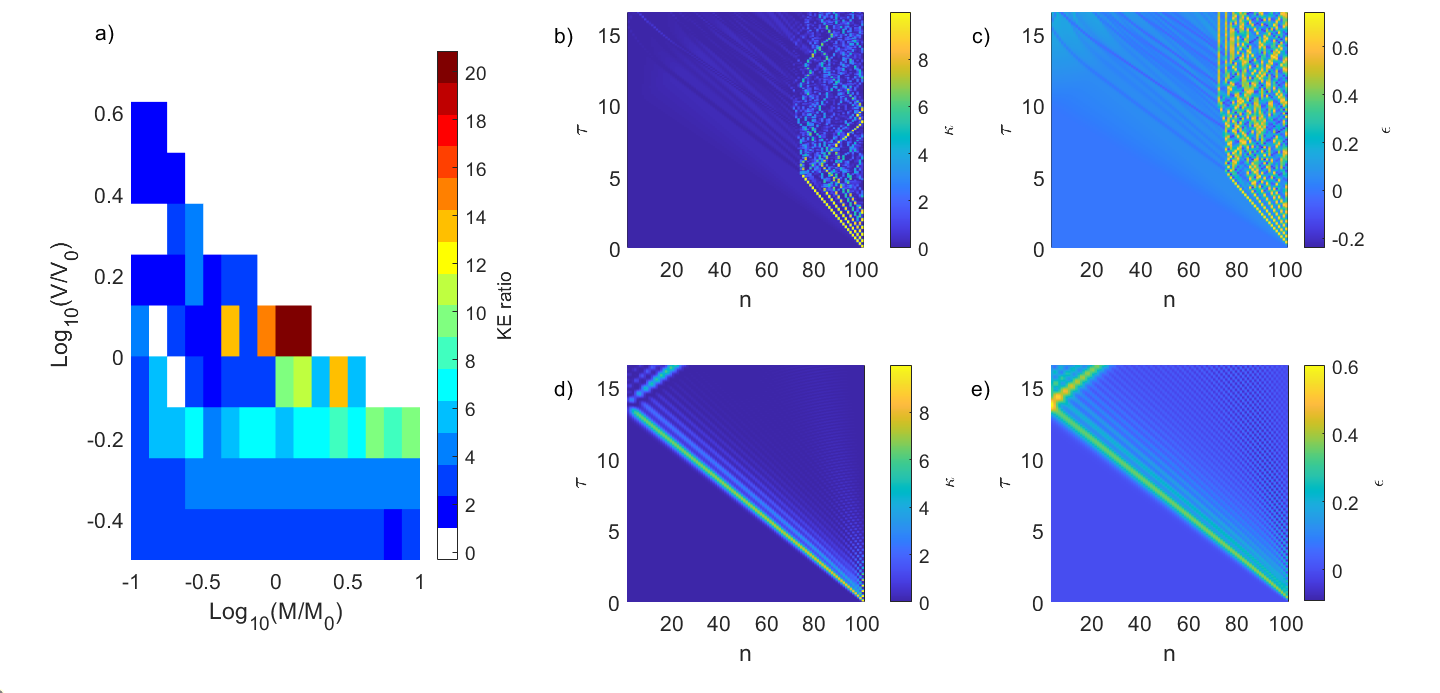}% 
\caption{Undamped simulations. (a) KE ratio as a function of impactor conditions. Non-dimensionalized KE, $\kappa$, for bistable (b) and linear (d) materials at nominal impact conditions ($M_0$, $V_0$). The colorbar is saturated at $\text{KE}_I/10$. Layer strain, $\epsilon$, for bistable (c) and linear (e) materials at nominal impact conditions ($M_0$, $V_0$).} 
\label{fig2}
\end{figure*}

\section{Simulated dependence on impact conditions for a conservative material}

To assess the performance of the bistable mechanical metamaterial, the linear ``control'' material was simulated using a linear stiffness corresponding to the slope between zero force and the force at $\varepsilon_0$ for the bistable material (equal to $1.77\beta_1$), identical mass, contact spring and damping parameters to the bistable sample. Simulations with varied impactor mass and velocity using an undamped linear material showed negligible difference in KE transmission at the $n=N/2$ (approximately $1\%$ maximum), between linear materials of different stiffnesses, when analyzed from impact time until the time the first wavefront hits the bottom of the sample. We thus define a ``KE ratio", as the maximum KE of the $N/2$ mass of the control lattice divided by the maximum KE of the $N/2$ mass of the bistable lattice, where a value greater than one indicates the bistable material is outperforming the linear material in terms of minimizing KE transmission. 

We initially simulated the response of $N=100$ layer conservative (\textit{i.e.} $\eta=0$) materials for duration $T_s=1.2\tau_l$ where $\tau_l$ is the time for the linear wave to travel across the material once (based on $c_0$). Figure \ref{fig2}(a) shows the KE ratio for varied $M$ and $V$, wherein the bistable material outperforms the linear material by up to 21x, while performing worse than the linear material particularly for impactor conditions where $M>M_0$ and $V>V_0$. A reduction in performance above the nominal impact conditions of $M_0$ and $V_0$ as they represent the point where the kinetic energy of the impactor could be equally distributed across the first half of the material and stably locked into strain energy. This is consistent with the approximate iso-energy diagonal threshold that can be observed in Fig.~\ref{fig2}(a) that separates KE ratios above and below unity. This however opens the question however, as to why the bistable material would perform less well than the linear material without any energy locking capability. To address this question we proceed to study the spatiotemporal response for both systems. 

To study the spatiotemporal response of the absorber materials at specific impact conditions, we define a normalized KE $\kappa = N (\text{KE}_n/\text{KE}_I)$, where $\text{KE}_n$ is the KE of the nth particle and $\text{KE}_I$ is the KE of the impactor, and strain $\epsilon=u/a$. The simulation time $t$ is expressed in terms of $\tau = t\sqrt{{\beta_1}/m}$. Figure \ref{fig2}(b,c) shows $\kappa$ and $\epsilon$ at impactor conditions $M_0$, $V_0$, and Fig. \ref{fig2}(d,e) shows the linear sample at the same impact conditions. We note that the impactor is not shown in the plots of $\epsilon$, although it is included as the $n=N+1$ layer in the $\kappa$ diagrams. 

Figure \ref{fig2}(d,e) shows a minimally dispersive pulse that propagates through the linear sample. In contrast, the bistable material (Fig. \ref{fig2}(b,c)) shows nonlinear effects, demonstrating the richness of the bistable system.  In Fig. \ref{fig2}(b,c), we see solitary, transition wave emission from the initial impact, as shown by the straight, yellow lines that progress $\sim 25$ particles into the sample from the point of impact. The identification of these as transition waves is noted by strains exceeding $\varepsilon_2$. Looking first at each of these emissions, we see that the first wave is the fastest, then the next two are progressively slower, which is consistent with the amplitude dependent wavespeed of many solitary waves \cite{dauxois2006physics}. Beyond $\tau\sim4$, between the $80-100$th particles, we see a complex oscillatory behavior, likely a combination of low amplitude vibrations and dynamic snapping and unsnapping of the bistable layers. Crucially, past $\sim 25$ particles from the impactor, very little KE is transmitted further into the material.

\begin{figure*}
\centering
  \includegraphics[width=\textwidth]{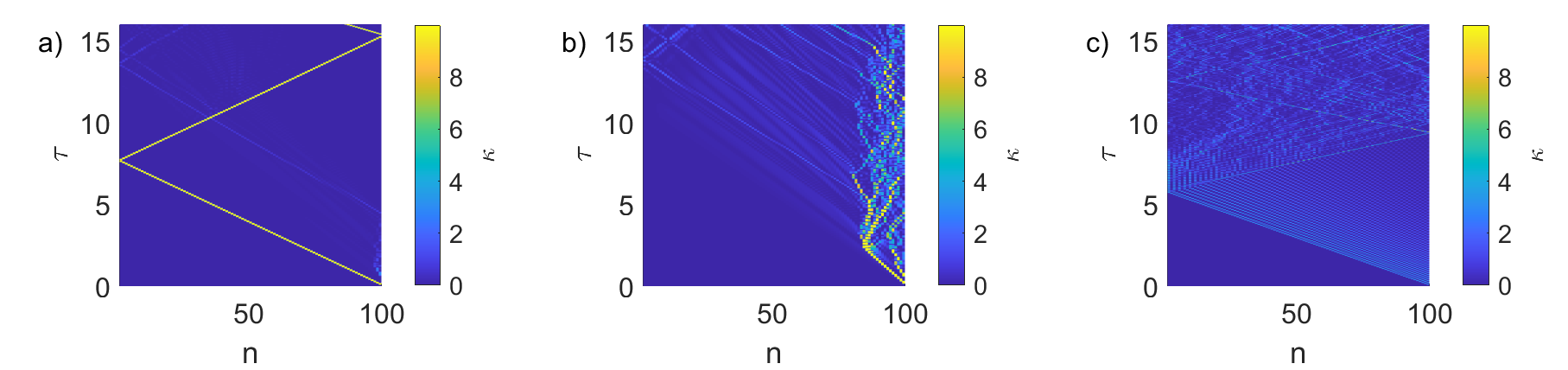}% 
\caption{Normalized KE, $\kappa$, for the simulated undamped bistable material shown in Fig.~\ref{fig2}(a) for the (a) worst performance (low mass and high velocity, $M/M_{0}=10^{-1}$ and $V/V_{0}=10^{0.75}$), (b) good performance (close to nominal mass and velocity, $M/M_{0}=10^{-0.625}$ and $V/V_{0}=10^{0.25}$), and (c) poor performance (high mass and high velocity, $M/M_{0}=10^{0.75}$ and $V/V_{0}=10^{0.75}$).} 
\label{Good_bad_ugly}
\end{figure*}

In order to investigate the changing performance shown in Fig.~\ref{fig2}(a), we study three different impactor conditions. Figure \ref{Good_bad_ugly}(a) shows a location to the top left of Fig. \ref{fig2}(a), which also corresponds to the worst performance of the sweep. At these impactor conditions ($M/M_{0}=10^{-1}$ and $V/V_{0}=10^{0.75}$), a high amplitude, short wavelength, transition wave travels throughout the sample, at speeds faster than the waves in the linear material, resulting in a KE ratio of $0.0848$x. Defining $M_{R}= M/m$, $M_{R}= 1.5$ and the formation of a single solitary wave in this instance agrees with previous studies, for instance in granular chain impacts \cite{nesterenko2013dynamics}, favoring single solitary wave formation when impactor and layer mass are closely matched. Within the context of solitary waves, nonlinear self-localization likely contributes to the poor performance seen in such cases. Figure \ref{Good_bad_ugly}(b) shows impactor conditions: $M/M_{0}=10^{-0.625}$ and $V/V_{0}=10^{0.25}$ representative of good performance (KE ratio $=4.44$x), but not as good as the nominal conditions shown in Fig. \ref{fig2}(b). A side by side comparison reveals qualitative differences in the behavior; namely longer lasting and more initial solitary waves generated in Fig. \ref{fig2}(b), which transition to an oscillatory phase encompassing more particles than Fig. \ref{Good_bad_ugly}(b). Additionally, close inspection at the $n=N/2$ location in Fig. \ref{Good_bad_ugly}(b) show wave-fronts that travel through particle $n=N/2$ of higher KE density than any of the crossing wave-fronts in Fig. \ref{fig2}(b). We observe that these wavefronts occur later in time than the time it would take for the first solitary wave to reach $n=N/2$ and appear to stem from dynamic unsnapping of the unit cells. Figure \ref{Good_bad_ugly}(c) shows an example for high $M$ and $V$ ($M/M_{0}=10^{0.75}$ and $V/V_{0}=10^{0.75}$), where $M_R =87$, and parallels could be drawn with known ``shock" cases in granular chains \cite{molinari2009stationary}). A ``train" of solitary waves is emitted after impact, and appear to continue to do so along the right side of Fig. \ref{Good_bad_ugly}(c). The KE density magnitude of subsequent emitted waves appears to decrease, and the highest KE density solitary wave, which is also the wave that triggers the maximum $\kappa$, is the first wave emitted from the time of initial impact. We note the fastest transition wavespeeds are shown under these high-energy impact conditions. The KE ratio was $0.246$x, which we note, does not occur at the first passage of the first transition wave but rather after the reflection of the first wave off the fixed bottom (after $\tau \approx 5$). Reflections and interference are representative of the complications of longer simulation duration in undamped simulations.

\section{Effect of damping on system response}

\begin{figure*}[h!]
\centering
  \includegraphics[width=\textwidth]{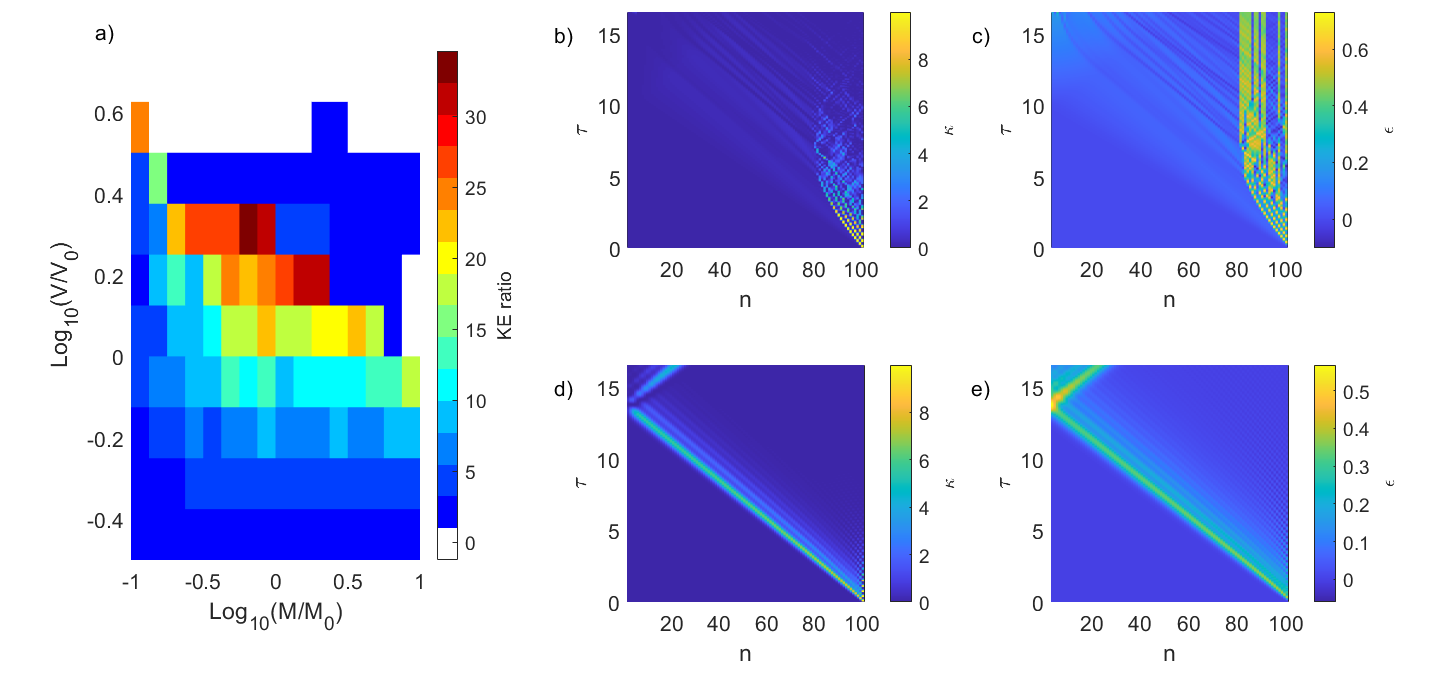}% 
\caption{Damped simulation. (a) KE ratio as a function of impactor conditions. Non-dimensional kinetic energy, $\kappa$, for bistable (b) and linear (d) materials at nominal impact conditions ($M_0$, $V_0$). The colorbar axis is saturated at $\text{KE}_I/10$. Layer strain, $\varepsilon$ for bistable (c) and linear materials (e) at nominal impact conditions ($M_0$, $V_0$).} 
\label{fig4}
\end{figure*}

The effect of the addition of a predetermined amount of damping to the simulations is assessed by implementing an intersite damping value of $\eta =0.164$ Ns/m, or a normalized value of ${\eta}{V_0}/{\beta_1}{\varepsilon_2}a=5.8e-3$ which represents the ratio of viscous to elastic effects. Otherwise, the simulation setup is identical to the prior section. Figure \ref{fig4} shows the results of the damped simulations in the same format as shown previously in Fig. \ref{fig2}. As before, Fig. \ref{fig4}(b,c) shows the bistable material response at impactor conditions $M_0$, $V_0$, and Fig. \ref{fig4}(d,e) shows the linear material response at the same conditions. The maximum KE ratio in the sweep in Fig. \ref{fig4}(a) was $34.58$x, which occurred at impactor conditions: $M/M_{0}=10^{-.25}$ and $V/V_{0}=10^{.25}$, noting a shift in the optimal impact conditions with the addition of damping. The maximum KE ratio from the sweep in Fig. \ref{fig4}(a) is higher than any values observed from the undamped sweep (Fig. \ref{fig2}(a)). In addition, Fig. \ref{fig4}(a) also shows more regions where the bistable material outperforms the linear material, as indicated by fewer regions of white color in the figure (noting the threshold of white to blue in the plot in Fig. \ref{fig2}(a) and Fig. \ref{fig4}(a) is a KE ratio of one).

Figure \ref{fig4}(b) shows four solitary waves that propagate in succession after impact, then transition to the oscillatory region by approximately $n=80$. In contrast to the undamped case, the wavespeed of each transition wave decreases after the point of impact, as could be expected by the damping-induced reduction of amplitude coupled with the amplitude dependent wavespeed of the solitary waves. Figure \ref{fig4}(c) shows regions with layers that have snapped to their secondary stable state and remained there, ostensibly aided by damping. Figure \ref{fig4}(d,e) show similar behavior of the damped linear sample as was seen in the undamped linear sample in Fig. \ref{fig2}(d,e), with the exception of a lower maximum $\kappa$ value at $n=N/2$ when damping is included.

\begin{figure*}[h!]
\centering
  \includegraphics[width=\textwidth]{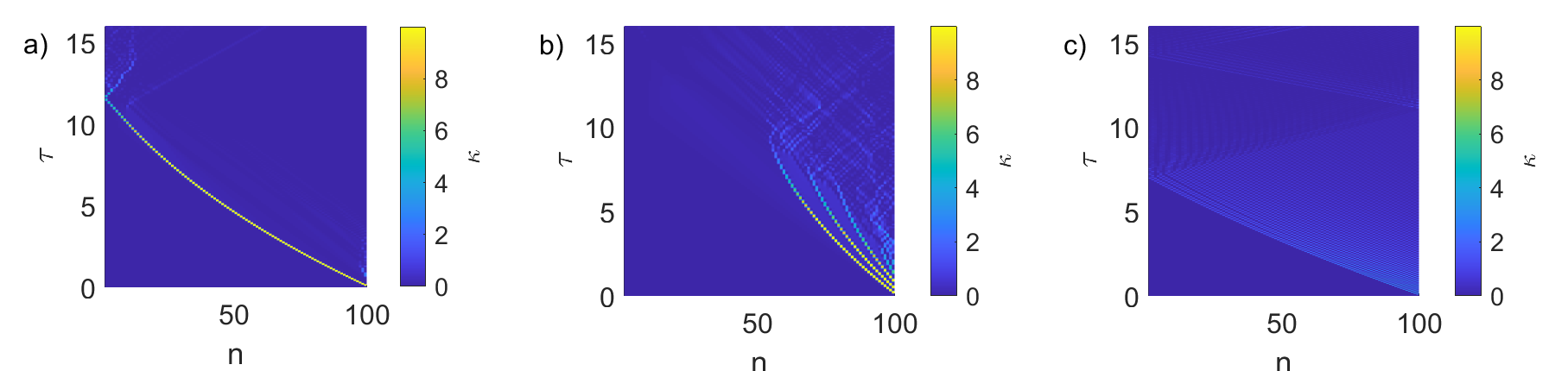}% 
\caption{Normalized KE, $\kappa$, for the simulated damped bistable material shown in Fig.~\ref{fig4}(a) for the (a) worst performance (low mass and high velocity, $M/M_{0}=10^{-1}$ and $V/V_{0}=10^{0.75}$), (b) best performance (close to nominal mass and velocity, $M/M_{0}=10^{-0.25}$ and $V/V_{0}=10^{0.25}$), and (c) poor performance (high mass and high velocity, $M/M_{0}=10^{0.75}$ and $V/V_{0}=10^{0.75}$).} 
\label{fig5}
\end{figure*}

Further comparison of $\kappa$ for three chosen impactor conditions picked from within Fig. \ref{fig4}(a) are shown in Fig. \ref{fig5}. Figure \ref{fig5}(a) shows impactor conditions corresponding to the top left corner and the worst performance within Fig. \ref{fig4}(a), representing impactor mass $M/M_{0}=10^{-1}$ and velocity $V/V_{0}=10^{0.75}$, and a KE ratio of $0.2073$x. It is also the same impactor conditions that resulted in the worst performance from the undamped sweep. Figure \ref{fig5}(b) shows the best performance, a KE ratio of $34.58$x, at impactor conditions $M/M_{0}=10^{-0.25}$ and $V/V_{0}=10^{0.25}$. Qualitatively, some similarities are observed when Fig. \ref{fig5}(b) is compared with the undamped high performance cases (Fig. \ref{fig2}(b) and Fig. \ref{Good_bad_ugly}(b)). We see an initial solitary wavefront that transitions (in time) to an oscillatory region prior to crossing particle $n=N/2$ (noting again the slowing wavespeed). Crucially, there appears to be an interaction between the solitary wave fronts and damping such that when multiple solitary wave emissions occur, it is possible that the waves slow as they reduce in amplitude (due to damping) such that they approach, but do not pass, the $n=N/2$ location. This may further be a synergistic effect in the sense that the damping-induced speed reduction gives the solitary waves even more time to decay in amplitude, and eventually vanish, before reaching the halfway point. Figure \ref{fig5}(c) shows impactor conditions $M/M_{0}=10^{0.75}$ and $V/V_{0}=10^{0.75}$, which is a region in the upper right corner of the sweep in Fig. \ref{fig4}(a), showing poor performance. A comparison of Fig. \ref{fig5}(c) to the undamped case (Fig. \ref{Good_bad_ugly}(c)) at the same impactor conditions shows similar behavior with the exception of the slowing transition waves in the damped case. 

\section{Conclusions}
The kinetic energy transmission performance in response to impact of a simulated bistable mechanical metamaterial was found to be highly dependent on the impactor conditions (mass and velocity). In the undamped simulations, the performance of a $N=100$ sample ranged from worse ($0.08$x) to far superior ($20.9$x) in comparison to the linear control lattice. Similarly, in the damped sweep, the performance again ranged from worse ($0.21$x) to far superior ($34.6$x). The presence of damping was seen to have a beneficial, potentially synergistic, effect on the performance of the system, as indicated by both higher maximum KE ratio values and higher minimum KE ratio values for the same set of impactor conditions. One likely reason for this is the higher amount of ``permanent" snapping of layers prior to the $n=N/2$ point. This permanent snapping is aided by damping both by the viscous resistance to unsnapping and reduction of traveling waves that may cause unsnapping at later times after subsequent reflections and constructive interference. 

Two significant additional implications of the findings regarding impact conditions and damping are the following. First, the bistable mechanism has the potential to yield significant performance benefits in terms of KE abatement if the material and impact conditions are well paired, but if not well paired, the bistable material can underperform a more traditional material, sometimes by a substantial margin. These high performance conditions were found, via simulation, to match well with predictions made using simple analytical estimates (namely, Eqs. \ref{v0eqn} and \ref{PEeqn}). Second, while only a single viscosity was used in the damped examples, the performance improvement as a result of the addition of damping leads the authors to suspect that subsequent ``tuning" of the damping may significantly affect the performance of these systems. 

We further suggest several potential avenues for future study in the context of both the fundamental understanding of energy transmission in bistable systems as well as their application in impact mitigation systems. The simulations considered herein idealized the response of a physically realized bistable mechanical metamaterial by not considering self contact (or full compaction) of the physical unit cells. The future inclusion of self contact will be critical for accurately describing the response of such systems, and can be expected to modify ranges of poor and superior impact performance. In addition, we suggest that computational optimization has a key future role to play, particularly given the strong nonlinearities involved herein. This may take the form of shape optimization of the unit cell aimed at maximizing energy absorption per unit mass density considering the metamaterial's dynamic response. One may also consider optimizing the unit cell to broaden the envelope of impactor conditions (\textit{e.g.} Fig. \ref{fig2} and Fig. \ref{fig4}) wherein the bistable material gives superior impact absorption response. Finally, we consider there may exist additional rich dynamics and unique capabilities for impact absorption with bistable media in higher dimensions, such as in the case of localized point impacts, instead of the 1D (plate-impact-like) scenarios considered herein.  

\section{Acknowledgments}

This project was supported by the US Army Research Laboratory and Army Research Office under grant no. W911NF-17-1-0595. I.F. acknowledges support from the Department of Defense (DoD) through the National Defense Science \& Engineering Graduate (NDSEG) Fellowship Program. B.M. acknowledges support from the U.S. Department of Energy (DOE) National Nuclear Security Administration (NNSA) Laboratory Graduate Residency Fellowship (LRGF) under Cooperative Agreement DE-NA0003960. The authors declare they have no competing financial interests.

%\bibliography{bistable_references}

\bibliographystyle{ieeetr}
\bibliography{references_2}

\end{document}